\documentstyle[preprint,aps]{revtex}
\def\sleq{\mathrel{\rlap {\lower 4pt\hbox {\hskip -1pt$\sim}}
\raise 1pt \hbox {$<$}}}
\begin{document}
\draft
\preprint{\vbox{Submitted to Physical Review C
                \hfill IU/NTC 94-11\\
                \null\hfill DOE/ER/40561-152-INT94-00-64}}

\title{Relativistic Models for Quasielastic $(e,e')$\ at Large
Momentum
Transfers.}

\author{ Hungchong Kim \footnote{ Email : hung@iucf.indiana.edu},
C. J. Horowitz \footnote{ Email : charlie@iucf.indiana.edu}}
\address{Nuclear Theory Center and Dept. of Physics,
Indiana University, Bloomington, IN 47408}
\author{M. R. Frank \footnote{ Email :
frank@ben.npl.washington.edu}}
\address{Institute for Nuclear Theory, University of Washington,
Seatle, WA
98195}
\maketitle
\begin{abstract}
Inclusive
quasielastic response functions are calculated for electron
scattering
in a relativistic model including momentum dependent scalar and
vector mean
fields.  The momentum dependence of the mean fields is taken from
Dirac optical
fits to proton nucleus scattering and is important in describing
data at
momentum transfers of 1 GeV/c or larger.  Our simple model is
applicable for
quasielastic scattering over a large range of momentum transfers.

\end{abstract}
\eject
\narrowtext
\section{INTRODUCTION}
\label{sec:intro}
There is growing interest in quasielastic electron scattering
experiments with
momentum transfers of order one GeV/c or larger.  Clearly these
require a
relativistic treatment.  Indeed, there have been many
relativistic mean field
and RPA calculations of quasielastic
scattering~\cite{kur,wehr,horo89}.
These calculations
often have large scalar $S$ and vector $V$ mean fields which
shift the mass of a
nucleon from $M$\ to $M^*$,
\begin{equation}
M^*=M+S\ .
\end{equation}
This changes the position of
the quasielastic peak from $\omega\approx {\bf q}^2/2M$\
(with ${\bf q}$\ the momentum and
$\omega$\ the energy transfer) to
\begin{equation}
$$\omega\approx \sqrt{{\bf q}^2+(M^*)^2} -
M^*\ .\label{qpeak}
\end{equation}
This change agrees well with the ``binding-energy" shift seen in
a variety of experiments at moderate momentum transfers
$| {\bf q} | \leq 600$\
MeV/c.

However, Eq.~(\ref{qpeak}) predicts very large shifts for
momentum transfers
of order
1000 MeV/c or larger.  Such large shifts are not seen in
data~\cite{chen}.
One would
like to understand this limitation and develop a simple
relativistic mean field
model that is applicable over a broad range of momentum
transfers.  Indeed,
most
of the relativistic models assume the self-energies or mean
fields ($S,V$) are
independent of energy or momentum.  This approximation is
probably good at low
momentum transfers but fails as the momentum transfer increases.

Dirac optical model fits to proton-nucleus elastic
scattering~\cite{bc1}
produce scalar
and vector optical potentials which have much less energy
dependence than the
nonrelativistic mean field.  When the Dirac equation is reduced
to a
Schrodinger
like form the effective nonrelativistic mean field is~\cite{bc}
\begin{equation}
U_{opt}=S + {E\over
M} V + {1\over 2 M} (S^2-V^2)\ .\label{pot}
\end{equation}
Here $E$ is the total energy of the
nucleon ($T_{lab}+M$).  The linear energy dependence for
$U_{opt}$\ implied by
Eq.~(\ref{pot}) agrees well with data for $T_{lab}$\ of order 200
to 300 MeV
or less.
However, Eq.~(\ref{pot}) (with constant $S,V$) greatly
overpredicts the energy
dependence of $U_{opt}$\ above 300 MeV.  As a result Dirac
optical fits need
{\it energy dependent} $S$\ and $V$\ in order to reproduce
elastic data at
higher energies.  One such fit is shown in Fig.~\ref{self}.

The magnitudes of $S$\ and $V$\ decrease with energy.  This trend
is also
reproduced in relativistic Brueckner
calculations~\cite{horo87,ansa}.
In this paper we use
such momentum or energy dependent potentials to calculate
inclusive quasielastic
electron scattering over a large range of momentum transfers.
Results for a
simplified form of momentum dependent potentials were presented
in an earlier
paper~\cite{frank}.  Our formalism is described in
Sec.~\ref{sec:for}.
Results for $^{56}$Fe
and $^{40}$Ca are given in Sec.~\ref{sec:results}.  This section
also
discusses current
conservation.  We summarize in Sec.~\ref{sec:sum}.

\section{FORMALISM}
\label{sec:for}

The inclusive $(e,e')$\ cross section for excitation energy
$\omega$\ and three
momentum transfer ${\bf q}$\ is
\begin{eqnarray}
{d^2\sigma\over d\Omega dE}=
\sigma_M\Bigl[{Q^4\over {\bf q}^4} R_L({\bf
q},\omega)+\Bigl({Q^2\over
2{\bf q}^2} +
{\rm tan}^2{\theta\over 2}\Bigr) R_T({\bf q},\omega)\Bigr]\
,\label{cross}
\end{eqnarray}
with the Mott
cross-section $\sigma_M=\alpha^2{\rm cos}^2(\theta/2)/4E^2{\rm
sin}^2(\theta/2)$.  Here $Q^2=- q_\mu^2={\bf q}^2-\omega^2$\
and $\theta$\ is the
scattering angle.  The longitudinal $R_L$\ and transverse $R_T$\
response
functions are calculated in a local density approximation from
nuclear matter at
a density $\rho=2 k_F^3/3\pi^2$,
\begin{eqnarray}
R_L&=&-{2\over \pi\rho} {\rm Im}
(Z\Pi_{00}^p+N\Pi_{00}^n)\label{long}\ ,\\
R_T&=&-{4\over \pi\rho} {\rm Im}
(Z\Pi_{22}^p+N\Pi_{22}^n)\ ,
\end{eqnarray}
for a target with $Z$\ protons and $N$\
neutrons.  (Note, {\bf q} is assumed to be along the ${\hat {\bf
1}}$
axis so the subscript
22 refers to a transverse direction.)

The polarization $\Pi$\ is calculated with the nucleon Greens
function $G(p)$\
and the electromagnetic vertex $\Gamma$,
\begin{equation}
\Pi_{\mu\nu}^i(q,\omega)= -i \int
{d^4p\over (2\pi)^4} {\rm Tr}[ G(p+q)\Gamma_\mu^i G(p)
\Gamma_\nu^i ]\ ,
\label{pol}
\end{equation}
for $i=$\ p (proton) or n (neutron).  We assume the
electromagnetic vertex has the
simple form,
\begin{equation}
\Gamma_\mu^i = F_1^i \gamma_\mu + F_2^i{i\sigma_{\mu\nu}q^\nu
\over 2M}\label{ver}
\end{equation}
even for off-shell nucleons in the medium.

We calculate $G$\ in a mean field approximation.  The nucleon
self energy
is assumed to be
\begin{equation}
\Sigma = \gamma_0 V({\bf p}) + S({\bf p})\ ,
\end{equation}
where $S$\ and $V$\
are taken from a Dirac optical fit at the self-consistent energy
$E_{\bf p}$,
\begin{equation}
E_{\bf p} = \sqrt{{\bf p}^2 + (M+S({\bf p}))^2} + V({\bf p})
\end{equation}
The Greens function for a Fermi
momentum $k_F$\ is
\begin{equation}
G(p)=(p^*_\mu \gamma^\mu + M^*_{\bf p})\Bigl[{1\over {p^*}^2-
{M^*_{\bf p}}^2 + i\epsilon} + {i\pi\over E^*_{\bf
p}}\delta(p_0-E_{\bf p})
\Theta(k_F-|{\bf p}|)\Bigr]\ .
\label{green}
\end{equation}
Here,
\begin{eqnarray}
p^*_\mu&=&\{ p_0-V({\bf p}),{\bf p}\}\ ,\\
M^*_{\bf p}&=&M+S({\bf p})\ ,
\end{eqnarray}
and $E^*_{\bf p}=\sqrt{{\bf p}^2+{M^*_{\bf p}}^2}$\ .

We use $S({\bf p})$\ and $V({\bf p})$\ from the real parts of an
optical
fit to p-$^{40}$Ca
elastic scattering.  This fit reproduces experimental data from
21 MeV up to a
laboratory kinetic energy of 1.04 GeV~\cite{bc1}.  For energies
below 21 MeV we
interpolate between this fit and the mean field approximation to
the Walecka
model values of~\cite{horo81}
\begin{eqnarray}
V_{MFT}&=&354\ {\rm MeV}\label{vector}\ ,\\
S_{MFT}&=&-431\ {\rm MeV}\label{scalar}\ .
\end{eqnarray}
For energies greater than 1.04 GeV we use the smooth
extrapolation shown in Fig.~\ref{self}.
(However, results we present are not sensitive
to this region.)  Note, the optical fit has imaginary parts for
$S$ and $V$
which we ignore.  It remains to investigate the effects of these
complex
potentials.

We use the fit potentials at the center of $^{40}$Ca.  However we
scale these
results by $\alpha$,
\begin{equation}
\alpha={\int d^3 r \rho(r)^2 \over \rho_0 \int d^3 r
\rho(r)}=(230/257)^3\ ,
\label{alp}
\end{equation}
to represent an average over the nuclear volume.  Here $\rho(r)$\
is taken from
a relativistic mean field calculation of $^{40}$Ca~\cite{horo81}.
We assume
the potentials scale with the density and
then use a Fermi momentum of
\begin{equation}
k_F=230\ {\rm MeV/c}\ ,
\end{equation}
in our
calculations rather than a typical nuclear matter value
$k_F^o\approx
257$\ MeV/c
with $\rho_0=2{k^o_F}^3/3\pi^2$.  This represents an average over
the central and
low density surface regions.

The polarizations in Eq.~(\ref{pol}) are calculated numerically
using the Greens
function in Eq.~(\ref{green}) as discussed in the appendix.  The
parameters of the
calculation are the Fermi momentum $k_F$, $\alpha$ and the two
functions
$S({\bf p})$\ and
$V({\bf p})$\ from the Dirac Optical fit.  This yields a simple
relativistic impulse
approximation that will be applied to data over a large range of
momentum
transfers in the next section.

\section{RESULTS}
\label{sec:results}

In this section we present results for three calculations.  The
first is a free
Fermi gas with $S({\bf p})=V({\bf p})=0$.  The second is the
Mean Field theory (MFT) with
momentum independent $V$\ and $S$\ given by
Eqs.~(\ref{vector}),~(\ref{scalar})
(scaled by the $\alpha$ of Eq.~(\ref{alp})).
Finally, we consider momentum dependent
$S({\bf p})$\ and $V({\bf p})$.

The longitudinal response for $^{56}$Fe at $|{\bf q}|=550$\ MeV/c
is shown in
Fig.~\ref{fe550}.
The free Fermi gas somewhat underestimates the average excitation energy
of the response.  This is better described by either the full momentum
dependent calculation or the
MFT.  Indeed at this relatively
low excitation energy of about 200 MeV there are not too large
differences between the momentum dependent $S({\bf p})$, $V({\bf p})$\ and
the MFT potentials.
Therefore, either the MFT or the momentum dependent calculation
can describe the average excitation energy of low momentum transfer data.

We have used a simple local density approximation.  This is
expected to reproduce qualitatively the position of the
quasielastic peak.  Full relativistic finite nucleus calculations
(for momentum independent potentials) ref.~\cite{horo89} agree
well with the local density peak positions and do a good job of
reproducing the detailed shape of the response.
Note, the area under the longitudinal response is controversial.
It is subject to systematic experimental errors and Coulomb
distortion corrections.  However,
there is general agreement that (near this momentum transfer)
there is a substantial binding energy shift compared to a free
Fermi gas.  This is seen either in separated longitudinal and
transverse responses or in unseparated
cross sections.

Next, Fig.~\ref{fe1140} shows the longitudinal response at a
momentum transfer of
$|{\bf q}|=1.14$\ GeV/c.  At this momentum transfer the MFT
response is at substantially
too high an excitation energy.  In contrast, either the momentum
dependent or the free Fermi gas calculation provides a reasonable
description of the position of the quasielastic peak.  Thus the
reduction of $S({\bf p})$\ and
$V({\bf p})$\ with increasing energy or momentum (as shown in
Fig.~\ref{self})
corrects (at least in large measure) the tendency for
the MFT to overpredict the binding energy shift.  At still higher
momentum transfers, we expect small differences between the
momentum dependent calculation and a free Fermi gas because
$S({\bf p})$\ and $V({\bf p})$\
continue to decrease.

The transverse response for $^{40}$Ca is shown in Fig.~\ref{tran}
for
$|{\bf q}|=$550
MeV/c.  Again, at this $q$\ the MFT and momentum dependent
calculations are similar.  Note, there is substantial extra
strength in the transverse response at high excitation energies
from meson
exchange currents and Delta production, see for
example~\cite{bruss}.
These are not contained in our simple model.

Note, a free Fermi gas reproduces the peak position well at high
$q$.  However, there is no theoretical motivation for simply
ignoring the mean fields which are known to be present.
Furthermore, a free Fermi gas will fail at low $q$\ where it is
generally believed one needs a binding energy shift.  Thus, we
think, the only theoretically consistent model is one which
includes mean fields and takes into account their momentum
dependence.

Alternatively, some nonrelativistic models simply assume a
momentum independent excitation energy shift equal to the average
binding energy of a nucleon.  This works reasonably well at low
$q$\ and possibly could be extended to high $q$\ by using some
form of relativistic kinematics.  However, there is no
theoretical justification for this prescription.  Strictly
speaking one would have to assume an energy dependent potential
which is big for bound states and then goes rapidly to zero for
positive energies.   This is unrealistic.
Thus, we believe, a theoretically consistent description will have
to address explicitly the momentum or energy dependence of the
mean fields.

We now discuss current conservation.  The full electromagnetic
vertex should
satisfy the Ward-Takahashi identity,
\begin{equation}
q_\mu\Gamma^\mu_{full} = G^{-1}(p+q) -
G^{-1}(p)\label{ward}\ .
\end{equation}
This is no longer true for the vertex $\Gamma^\mu$, of
Eq.~(\ref{ver}) given
that $G$\ is calculated with momentum dependent potentials (and
that
Eq.~(\ref{ver})
has form factors).  Clearly we must add a vertex correction.
However, there is
no unique way to calculate this correction without detailed
knowledge of the
processes included in
$S({\bf p})$ and $V({\bf p})$ and in $F_1$ and $F_2$.

Instead, we add a very minimal vertex correction so that not
Eq.~(\ref{ward})
but
at least current conservation is satisfied,
\begin{equation}
\bar U(S({\bf p}),V({\bf p})) q_\mu
\Gamma_{full}^\mu U(S({\bf p}),V({\bf p})) = 0\ .
\end{equation}
Here, the spinors $U$\ satisfy a
Dirac equation with $S,V$\ and
\begin{equation}
\Gamma_{full}^\mu = \Gamma^\mu +
\Delta\Gamma^\mu \ .\label{gam}
\end{equation}
We choose,
\begin{equation}
{\Delta\Gamma^\mu}^i = q^\mu
{F_1^i(q_\lambda^2)\over q_\lambda^2}\bigl[ S({\bf p+q})+\gamma_0
V({\bf p+q})-S({\bf p})-\gamma_0 V({\bf p})\bigr]\ .\label{dgam}
\end{equation}
This simple prescription
insures,
\begin{equation}
q^\mu \Pi_{\mu\nu}=\Pi_{\mu\nu}q^\nu=0\ .
\end{equation}
Calculations using
Eqs.~(\ref{gam}), (\ref{dgam}) are shown in
Figs.~\ref{fe550}, \ref{fe1140} and are only slightly smaller
than
calculations with $\Delta\Gamma^\mu=0$ \footnote{
Note, Eq.~(\ref{dgam}) does not contribute to $R_L$\ when dotted
into
the conserved
electron current.  However, Eqs.~(\ref{cross}), (\ref{long})
assumed current
conservation.  One gets
a different answer for $R_L$\ from Eq.~(\ref{long}) if one sets
$\Delta\Gamma_\mu=0$\
because in this case the current is not conserved.}.  Therefore,
we do not
expect large
errors from our (slight) violation of current conservation.
However,
Eq.~(\ref{dgam})
is non-unique and this point deserves further study, see below.

\section{SUMMARY AND CONCLUSIONS}
\label{sec:sum}

There have been many relativistic mean field calculations for
inclusive quasielastic electron scattering.  Ironically, although
these do a reasonable job at low to moderate momentum transfers,
many of these relativistic calculations are not appropriate for
momentum transfers near one GeV/c or above.
This is because they assume the mean fields are independent of
momentum or energy.  As a result they predict too large a binding
energy shift in the position of the quasielastic peak.

Instead, we assume a simple model where the momentum dependence
of the
scalar and vector mean fields are taken from an optical model fit
to p-$^{40}$Ca
elastic scattering.  The inclusive response is calculated in a
relativistic
impulse approximation.  Reasonable predictions are made for the
position of the
quasielastic peak at both low and high ${\bf q}$.  Therefore, the
model can be used over a large range of momentum transfers.

Our calculations are for nuclear matter in a local density
approximation.
We expect this to qualitatively reproduce the position of the
quasielastic peak
but not its detailed shape.  One should repeat the calculations
in a full finite
nucleus formalism, such as from reference~\cite{horo89}, using
energy dependent potentials.

We have only briefly mentioned vertex corrections.  These are
needed for current
conservation in momentum dependent mean fields.  Future work
should calculate
the vertex correction in a full microscopic model.  For example,
one can choose a
(somewhat unusual) set of meson couplings (including nonlinear
self-couplings)
so that a relativistic Hartree-Fock (HF) calculation
approximately reproduces
the momentum dependence of the mean fields.  This allows the
vertex correction
to be calculated explicitly from the diagram in
Fig.~\ref{vertex}.

\acknowledgments

We would like to thank B. C. Clark for providing us a computer
code
``global'' which generates the nuclear optical potentials.
This research was supported by U.S. Department of Energy under
Grant Nos.\ DE-FG02-87ER-40365 and DE-FG06-90ER-40561.

\eject
\narrowtext
\appendix
\section*{}

Here we summarize the calculation of imaginary part of the
polarization
$\Pi^{\mu\nu}$.   The Greens function of Eq.~(\ref{green}) can be
written
in three parts;
(1) the propagation of an antinucleon in the Dirac sea, the
propagation of a (2)
hole
inside and a (3) particle outside of the Fermi sea.
Since vacuum polarization does not contribute to the
imaginary part for the space-like $q_\mu$,  the
propagation of antinucleons can be dropped and the Greens
function
Eq.~(\ref{green}) written,
\begin{equation}
G(p) = {p^*_\mu \gamma^\mu + M^*_{\bf p} \over 2 E^*_{\bf p}}\
\Biggr[{\Theta (k_F
-|{\bf p}|) \over p_0-E_{\bf p}-i \epsilon}
+{\Theta (|{\bf p}|-k_F)
\over p_0-E_{\bf p}+i\epsilon}\Biggl]\ .
\end{equation}
Using this Greens function, it is straight forward to integrate
over $p_0$
and the imaginary
part of the polarization becomes,
\begin{eqnarray}
{\rm Im}\ \Pi^{\mu\nu} &&= -\int {{\bf p}^2 d|{\bf p}|\,d{\rm
cos}\theta
\over 4
\pi E^*_{\bf p}
E^*_{\bf p+q}} \Theta(|{\bf p+q}|-k_F)\ \Theta(k_F-|{\bf p}|)\
\delta(
E_{\bf p+q}-E_{\bf p}-q_0) \nonumber \\
&&\times [T_1^{\mu\nu}+T_2^{\mu\nu}+T_3^{\mu\nu}]\ ,
\label{imag}
\end{eqnarray}
where ${\rm cos}\theta$ is the angle between ${\bf p}$ and ${\bf
q}$.
Here $T_1, T_2$ and $T_3$ are from the trace of Dirac matrices,
\begin{eqnarray}
T^{\mu\nu}_1&=&{F_1^i}^2\ [(p+q)^{* \mu} p^{* \nu}-p^*\cdot
(p+q)^*
g^{\mu\nu} +p^{* \mu}
(p+q)^{* \nu} +M^*_{\bf p} M^*_{\bf p+q} g^{\mu\nu}]\ ,\\
T^{\mu\nu}_2&=&{F_1^i\ F_2^i\over 2 M} \biggr[ M^*_{\bf p}\
[(p+q)^*\cdot q\ g^{\mu\nu}-
(p+q)^{* \mu} q^\nu]+M^*_{\bf p+q}\
[p^{*\mu}q^\nu-p^*\cdot q g^{\mu\nu}]\biggl]\ ,\\
T^{\mu\nu}_3&=&{{F_2^i}^2 \over 4 M^2}\Biggr[[M^*_{\bf p+q}
M^*_{\bf p}+
p^*\cdot
(p+q)^*]\ (g^{\mu\nu}\ q_\mu^2-q^\mu q^\nu) \nonumber \\
&&+p^*\cdot q\ [(p+q)^{* \mu} q^\nu+(p+q)^{* \nu} q^\mu]
-q_\mu^2\ [(p+q)^{* \mu} p^{*\nu}+(p+q)^{* \nu} p^{*\mu}]
\nonumber \\
&&+(p+q)^*\cdot q\ [
q^\mu p^{*\nu}+q^\nu p^{*\mu}]-2 p^*\cdot q\ (p+q)^*\cdot q\
g^{\mu\nu}\Biggl]
\end{eqnarray}
with $p^{*\mu}=\{E_{\bf p}-V({\bf p}), {\bf p}\}$.  Equation
(\ref{imag}) is
evaluated
numerically. The
phenomenological self-energies $(V({\bf p}), S({\bf p}))$ are
fitted to
polynomials in ${\bf p}$.   Finally, the angle integration of
the $\delta$-function restricts the momentum integration
so that $|{\rm cos}\theta| \le 1$.
The form factors $F^i_1$ and $F^i_2$ are taken from
Ref.~\cite{don92b}.

\begin{figure}
\caption{Scalar $S$ and vector $V$\ mean fields versus laboratory
kinetic energy $T_{lab}$.  Solid lines are real parts of a Dirac
optical
model fit to p-$^{40}$Ca elastic scattering \
\protect{\cite{bc1}}
while dot-dashed lines are the
constant relativistic mean field values\ \protect{\cite{horo81}}.
The dashed
lines are extrapolations as described in the text.}
\label{self}
\end{figure}

\begin{figure}
\caption{Longitudinal response function for $^{56}$Fe at a
momentum transfer
of q=550 MeV versus excitation energy $\omega$.  The solid curve
is the full
momentum dependent calculation including the vertex correction
from
Eq.~(\protect{\ref{dgam}})
while the dot-dashed curve omits this vertex correction.  The
short-dashed
curve is
the response for a free Fermi gas while the dashed curve is the
response
assuming constant relativistic mean field theory self-energies.
The data is from Ref.\ \protect{\cite{mez}}.}
\label{fe550}
\end{figure}

\begin{figure}
\caption{Longitudinal response for $^{56}$Fe at a momentum
transfer of q=1.14
GeV/c.  See the caption to Fig.~\ \protect{\ref{fe550}}.
The data is from Ref.\ \protect{\cite{chen}}.}
\label{fe1140}
\end{figure}

\begin{figure}
\caption{Transverse response for $^{40}$Ca at a momentum transfer
of q=550 MeV/c.
The solid curve is the full momentum dependent calculation while
the short-dashed is a
free Fermi gas and the dashed is for constant relativistic mean
field theory
self energies.  The data is from Ref.\ \protect{\cite{barr}}}
\label{tran}
\end{figure}

\begin{figure}
\caption{Vertex correction in a relativistic Hartree-Fock
approximation.
The nucleon line is solid while the dashed line is an exchanged
meson and the
wavy line a photon.} \label{vertex}
\end{figure}


\begin{references}
\bibitem{kur}  {H. Kurasaw and T. Suzuki,
               Nucl.\ Phys.\ {\bf A490}, 571 (1988).}
\bibitem{wehr}  {K. Wehrberger and F. Beck,
               Phys.\ Rev.\ C {\bf 37}, 1148 (1988).}
\bibitem{horo89}  {C. J. Horowitz and J. Piekarewicz,
                Phys.\ Rev.\ Lett.  {\bf 62}, 391 (1989);
                Nucl.\ Phys.\ {\bf A511}, 461 (1990).}
\bibitem{chen} {J. P. Chen {\it et al.}
                Phys.\ Rev.\ Lett. {\bf 66}, 1283 (1991).}
\bibitem{bc1}{E. D. Cooper, S. Hama, B. C. Clark and R. L.
Mercer,
             Phys.\ Rev.\ C {\bf 47}  297 (1993).}
\bibitem{bc}  {B. C. Clark, S. Hama and R. L. Mercer in
              {\it The Interaction Between Medium Energy Nucleons
in Nuclei},
               edited by H. O. Meyer, AIP 1983.}
\bibitem{horo87}  {C. J. Horowitz and B. D. Serot,
                   Nucl.\ Phys.\ {\bf A464}, 613 (1987).}
\bibitem{ansa}  {M. R. Ansatasio, L. S. Celenza, W. S. Pong and
C. M. Shakin,
                 Phys.\ Rep. {\bf 100}, 327 (1983);
                 R. Machleidt, in
               {\it Advances in Nuclear Physics},
               edited by J.W. Negele and E. Vogt
               (Plenum, New York, 189 1986), Vol. 19.;
                 G. Q. Li, R. Machleidt and Y. Z. Zhuo
                 Phys.\ Rev.\ C {\bf 48}, 1062 (1993);
                 B. Ter Haar and R. Malfliet,
                 Phys.\ Rep. {\bf 149}, 207 (1987).}

\bibitem{frank} {M. R. Frank,
                 Phys.\ Rev. \ C {\bf 49}, 555 (1994).}
\bibitem{horo81}  {C. J. Horowitz and B. D. Serot,
                   Nucl.\ Phys.\ {\bf A368}, 503 (1981).}
\bibitem{bruss}  {M. J. Dekker, P. J. Brussaard and J. A. Tjon,
                  Phys.\ Lett. {\bf 289B}, 255 (1992);
                  M. J. Dekker, P. J. Brussaard and J. A. Tjon
                  Phys.\ Rev. C {\bf 49}, 2650 (1994)}
\bibitem{barr}  {P. Barreau {\it et al.},
               Nucl.\ Phys.\ {\bf A402}, 515 (1983).}
\bibitem{mez} {Z. E. Meziani {\it et al.}
                Phys.\ Rev.\ Lett. {\bf 52}, 2130 (1984).}
\bibitem{don92b}{M. J. Musolf and T. W. Donnelly,
                Nucl.\ Phys.\ {\bf A546}, 509 (1992).}
\end{references}
\end{document}